%% file: main.tex
\documentclass[runningheads]{llncs}
\usepackage[T1]{fontenc}
\usepackage{graphicx,verbatim}
\usepackage{authblk}
\usepackage{graphicx}
\usepackage{subfigure}
\usepackage{array}
\usepackage{float}
\usepackage{booktabs}
\usepackage{bbding}
\usepackage{etoolbox}
\usepackage{color}
\usepackage{soul}
\usepackage{algorithm}
\usepackage{algorithmic}
\usepackage{tabularx}
\usepackage{cite}
\usepackage{balance}
\usepackage{amsmath, amssymb}
\usepackage{makecell}
\usepackage{bm}
\usepackage{multirow}
\usepackage{color,xcolor}
\usepackage{hyperref}
\usepackage{colortbl}
\usepackage{orcidlink}

\hypersetup{
    colorlinks,
    linkcolor={blue!50!black},
    citecolor={blue!50!black},
    urlcolor={blue!50!black}
}

\usepackage[T1]{fontenc}
\begin{document}
\title{SurgTPGS: Semantic 3D Surgical Scene Understanding with Text Promptable  Gaussian Splatting}
%
%
\titlerunning{SurgTPGS}
\authorrunning{Y. Huang et al.}
\author{Yiming Huang\inst{1,2~\star}
\and Long Bai\inst{1,2,3~\star}
\and Beilei Cui\inst{1,2}
\thanks{Co-first authors.}
\and Kun Yuan\inst{3, 4}
\and Guankun Wang\inst{1, 2}
\and Mobarak I. Hoque\inst{5}
\and Nicolas Padoy\inst{4}
\and Nassir Navab\inst{3}
\and Hongliang Ren\inst{1,2}
\thanks{Corresponding author.}}
\institute{The Chinese University of Hong Kong, Hong Kong SAR, China
\and Shenzhen Research Institute, CUHK, Shenzhen, China
\and Technical University of Munich, Munich, Germany 
\and University of Strasbourg \& IHU Strasbourg, Strasbourg, France 
\and University College London, London, United Kingdom
\\
\email{\{yhuangdl, b.long, beileicui\}@link.cuhk.edu.hk, hren@cuhk.edu.hk}}


\maketitle              
\begin{abstract}

\input{sections/0_abstract}
\end{abstract}
\section{Introduction}
\input{sections/1_introduction}
\section{Methodology}
\input{sections/2_methodology}
\section{Experiments}
\input{sections/3_experiments}
\section{Conclusions}
\input{sections/4_conclusion}

%
%
%
%
\begin{credits}
\subsubsection{Acknowledgements.}
This work was supported by Hong Kong RGC CRF C4026-21G,  RIF R4020-22, GRF 14211420, 14216020 \& 14203323).

\subsubsection{\discintname} The authors have no competing interests to declare that are
relevant to the content of this article.
\end{credits}

\bibliographystyle{splncs04}
\bibliography{reference}
%




\end{document}

%% file: sections/0_abstract.tex
In contemporary surgical research and practice, accurately comprehending 3D surgical scenes with text-promptable capabilities is particularly crucial for surgical planning and real-time intra-operative guidance, where precisely identifying and interacting with surgical tools and anatomical structures is paramount. However, existing works focus on surgical vision-language model (VLM), 3D reconstruction, and segmentation separately, lacking support for real-time text-promptable 3D queries. In this paper, we present SurgTPGS, a novel text-promptable Gaussian Splatting method to fill this gap. We introduce a 3D semantics feature learning strategy incorporating the Segment Anything model and state-of-the-art vision-language models. We extract the segmented language features for 3D surgical scene reconstruction, enabling a more in-depth understanding of the complex surgical environment. We also propose semantic-aware deformation tracking to capture the seamless deformation of semantic features, providing a more precise reconstruction for both texture and semantic features. Furthermore, we present semantic region-aware optimization, which utilizes regional-based semantic information to supervise the training, particularly promoting the reconstruction quality and semantic smoothness. We conduct comprehensive experiments on two real-world surgical datasets to demonstrate the superiority of SurgTPGS over state-of-the-art methods, highlighting its potential to revolutionize surgical practices. Our code is available at:~\url{https://github.com/lastbasket/SurgTPGS}.

\keywords{ 3D Scene Understanding \and  Text-Promptable Segmentation \and Robotic Surgery.}

%% file: sections/1_introduction.tex
In modern computer-assisted interventions (CAI) procedures, accurately understanding 3D surgical scenes has become a cornerstone for successful surgical outcomes~\cite{mclachlan20112d}. Previous studies~\cite{allan20202018, hong2020cholecseg8k} have emphasized the significance of incorporating semantic segmentation as pixel-wise surgical guidance, providing context awareness of tissues and instruments for the surgeon~\cite{jin2022exploring,yoon2022surgical,liu2023lskanet}. Despite significant progress in related fields, the vision-language 3D understanding of surgical scenes remains largely unexplored. This knowledge gap critically hampers semantic-aware planning and navigation in robot-assisted surgery. Without a comprehensive vision-language 3D understanding, robotic systems struggle to interpret the surgical environment accurately.

Recent breakthroughs in neural rendering~\cite{mildenhall2021nerf, kerbl20233d} enable high-quality reconstruction and real-time rendering, revolutionizing the surgical scene reconstruction and rendering~\cite{wang2022neural,zha2023endosurf,huang2024endo,liu2024endogaussian,zhu2024deformable,huang2025advancing}. Subsequent  extensions~\cite{kerr2023lerf,qin2024langsplat, wu2024opengaussian} enable the 3D-language interaction by incorporating the vision-language model (VLM)~\cite{radford2021learning}. 
In surgical understanding, VLMs~\cite{seenivasan2022surgical,yuan2023learning} provide textual guidance but are limited to 2D analysis, while visual question localized-answering (VQLA)~\cite{bai2023surgical,li2024llava} adds anatomical grounding. The recent works~\cite{zhou2023text, wang2024video} integrate the vision-language model with surgical tool segmentation, achieving diverse surgical instruments in minimally invasive surgeries. However, current 3D surgical scene reconstruction methods~\cite{yang2024deform3dgs, huang2024endo, liu2024endogaussian} lack language-aligned anatomical understanding, critically limiting their ability to interpret clinician intents. While surgical VLM~\cite{seenivasan2022surgical,bai2023surgical} and segmentation networks~\cite{jin2022exploring,yoon2022surgical,liu2023lskanet} remain confined to 2D frames, failing at the spatial level in 3D scenes.  This deficiency means they cannot meet the real-time demands of intraoperative navigation, where surgeons need immediate access to accurate 3D information to make informed decisions during procedures.

\begin{figure}[!t]
    \centering
    \includegraphics[width=\linewidth]{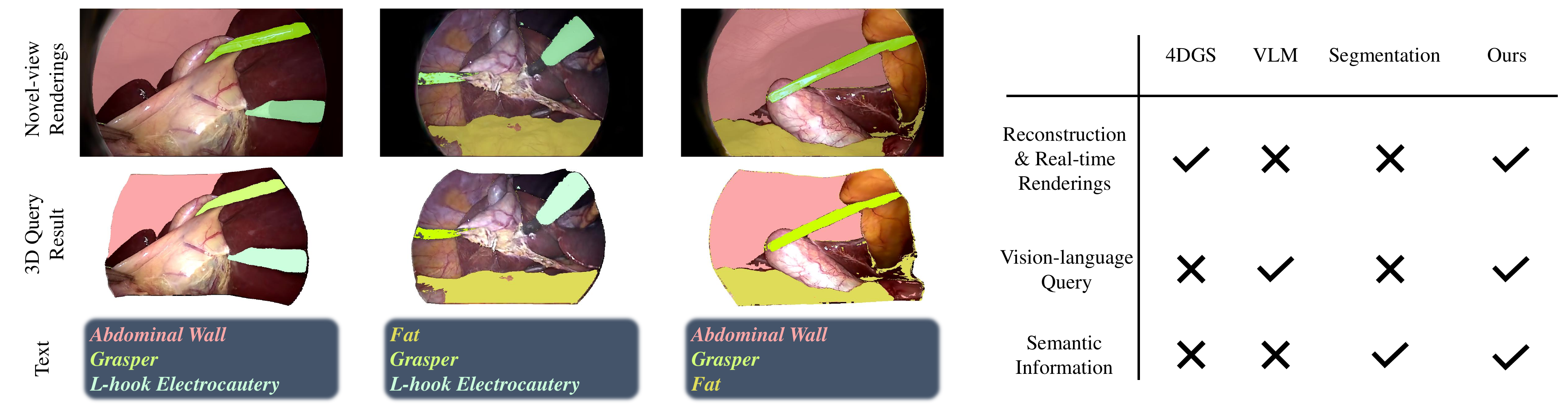}
    \caption{SurgTPGS provides 3D semantic-segmentation with text-promptable queries, complementing the drawbacks of 4DGS, VLM, and segmentation networks.}
    \label{fig:intro}
\end{figure}

 Although general-domain 3D-language models~\cite{kerr2023lerf,qin2024langsplat,wu2024opengaussian, labe2024dgd} achieve high accuracy in semantic queries for rigid objects, they fail in surgical scenes due to bleeding, occlusion, and soft-tissue motion. Besides, surgical anatomy's dynamic and deformable nature introduces unique challenges for semantic Gaussian training, causing inconsistent feature alignment during the training progress. This inconsistent alignment disrupts the spatial coherence of the appearance and the semantic feature for the 3D Gaussian representation, leading to blurred and corrupted training results. The sub-optimal semantic features cause the mismatch between vision and textual embedding, degrading the final query accuracy. To complement this research gap, we introduce SurgTPGS, a novel text-promptable Gaussian Splatting for 3D surgical semantic query. We present the \textit{3D semantics feature learning} strategy, integrating the Segment Anything model (SAM)~\cite{kirillov2023segment} with vision-language surgery features. To account for the dynamic nature of surgical scenes, we propose \textit{semantic-aware deformation tracking}, capturing the semantic-level deformation for more high-quality and precise reconstruction for both appearance and semantic features.  Furthermore, we introduce \textit{semantic region-aware optimization} with regional-based semantic supervision, improving the reconstruction quality semantic smoothness for object identification.
As shown in Fig~\ref{fig:intro}, SurgTPGS could offer surgeons real-time, semantic-aware assistance with language queries, improving surgical outcomes and patient care. Specifically, the contributions of this paper are as follows:
\begin{itemize}
    \item We present SurgTPGS, the \textbf{first} text-promptable Gaussian Splatting approach for 3D surgical scene understanding.
    \item We introduce a 3D semantic feature learning strategy integrating the SAM with VLM for in-depth surgical scene understanding.
    \item We propose semantic-aware deformation tracking to capture semantic deformation, enabling more precise texture and semantic feature reconstruction. 
    \item We propose semantic region-aware optimization, which uses regional-based semantic information to supervise training and improve reconstruction quality and semantic smoothness.
    \item We show that SurgTPGS achieves state-of-the-art results through comprehensive experiments on real-world surgical datasets.
\end{itemize}

%% file: sections/2_methodology.tex
\begin{figure}[!t]
    \centering
    \includegraphics[width=\linewidth]{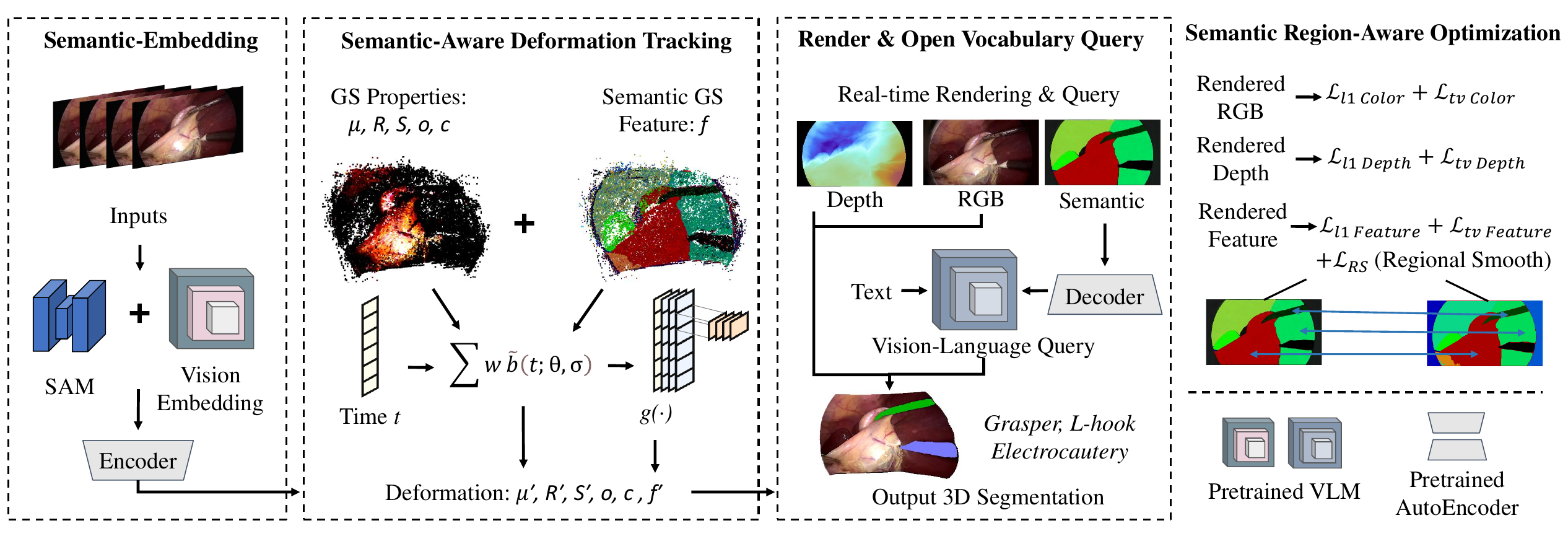}
    \caption{\textbf{Overview of SurgTPGS.} We first extract semantic embedding from SAM~\cite{kirillov2023segment} and VLM~\cite{cho2024cat}; then, we train the deformable Gaussians with semantic-aware deformation tracking and semantic-region-aware optimization. SurgTPGS supports real-time semantic 3D query and novel-view rendering simultaneously.}
    \label{fig:main}
\end{figure}

As illustrated in Fig~\ref{fig:main}, our method first extracts semantic embedding with SAM~\cite{kirillov2023segment} and CAT-Seg~\cite{cho2024cat}, then we train the deformable Gaussian with the semantic features by \textit{semantic-aware deformation tracking} and \textit{semantic region-aware optimization}. We achieve text-promptable 3D query by mapping the rendered semantic features with the text embedding.

\subsection{Preliminaries}\label{sec.2.1}

3D Gaussian Splatting (3DGS) represents scenes using a 3D Gaussian point cloud with learnable attributes, enabling efficient real-time rendering and inspiring various applications. The 3DGS is represented as $\mathcal{G} = \{\mu, r, s, o, \mathbf{SH}\}$, where $\mu,r,s,o,\mathbf{SH}$ are the mean, rotation, scale, opacity and spherical harmonic for color representation. To apply 3DGS in surgical scene reconstruction, Deform3DGS~\cite{yang2024deform3dgs} proposed the flexible deformation modeling scheme (FDM) to deal with deformable tissue challenges:
\begin{equation}
    \psi\left(t ; \Theta\right)=\sum_{j = 1}^{B}\omega_{j} \tilde{b}\left(t ; \theta_{j}, \sigma_{j}\right), \ | \ \tilde{b}(t ; \theta, \sigma)=\exp\left(-\frac{1}{2\sigma^{2}}(t - \theta)^{2}\right),
\end{equation}
where $\tilde{b}(\cdot)$ is the Fourier and polynomial basis function, $\{\omega, \theta, \sigma\}\in \Theta$ are learnable weight, center, and variance. For mean, rotation, and scale, the deformation is represented as $\mu^\prime=\mu+\psi(t;\Theta^\mu)$, $r^\prime=r+\psi(t;\Theta^r)$, and $s^\prime=s+\psi(t;\Theta^s)$. The mean and covariance~\cite{zwicker2001surface} are projected into 2D with the given camera parameters and rendered into color $C$ and depth $D$ with alpha blending~\cite{kerbl20233d}.

\subsection{Proposed Methodology}\label{sec.2.2}

\noindent \textbf{3D Semantic Feature Learning in Surgical Scenes.}
Inspired by~\cite{qin2024langsplat}, we introduce 3D Semantics feature learning with SAM~\cite{kirillov2023segment} as the segmentation generator and CAT-Seg~\cite{cho2024cat} as our vision-language baseline. We use SAM to obtain a segmentation mask $M$ and then gather the vision-language feature from CAT-Seg within the segmented areas. Similar to~\cite{qin2024langsplat}, we train an autoencoder $\mathbf{\Phi}$ for compressing the feature dimension for more efficient training. For pixel $x$, the pixel-aligned semantic feature can be obtained as follows:
\begin{equation}
    \hat{F}(x)=\mathbf{\Phi}_{encode}(V_{CAT - Seg}\left(I \odot M(x)\right),
\end{equation}
where $I$ is the input image, $\odot$ is the element-wise multiplication, $\mathbf{\Phi}_{encode}$ is the encoding function, and $V_{CAT-Seg}$ is the feature extraction function of CAT-Seg.\\

\noindent \textbf{Semantic-Aware Deformation Tracking.}
 Instead of using the deformation model as the mean, rotation, and scale from~\cite{yang2024deform3dgs}, our proposed method utilizes a convolution-based network to enhance the spatial consistency for semantic feature representation. Building on the semantic features obtained from our 3D semantics feature learning, we track the deformation of semantic regions with a lightweight spatial-aware network ${g}=(\text{Conv1d}(\cdot))$. For a feature $f$ associate with Gaussian, the deformed $f^\prime$ is computed as:
 \begin{equation}
     f^\prime = f+{g}(\text{Concat}(f,t))\cdot\beta, \ | \ \beta=1/(1+\exp(-\delta\psi(t; \Theta^f))),
 \end{equation}
 where $\beta$ is the probability map from the sloped sigmoid. $t$ is the time, and $\delta$ is the sigmoid slope, where 2.5 is applied. Similar to the color rendering~\cite{kerbl20233d}, the final semantic feature map $F$ is rendered from $f^\prime$ using alpha blending. With the proposed semantic-aware deformation tracking, we achieve spatially consistent modeling of semantic feature deformation.
\\

\noindent \textbf{Text-Promptable Surgical Scene Querying.}
By obtaining the pixel-aligned semantic feature $F$ from rendering, we query the 3D language feature by the language embedding $\phi_{text}$. The rendered feature is first decoded back into the image embedding $\phi_{img}=\mathbf{\Phi}_{decode}(F)$, and then following~\cite{kerr2023lerf}, we calculate the relevancy score as:
\begin{equation}
    score = {\min}_i \frac{\exp(\phi_{img}\cdot \phi_{text})}{\exp(\phi_{img}\cdot \phi_{text})+\exp(\phi_{img}\cdot \phi^i_{canon})},
\end{equation}
where $\phi_{canon}^i$ is a set of CLIP embeddings from the canonical phrase \textit{"object"}, \textit{"things"}, \textit{"stuff"}, and \textit{"texture"}. After computing the score, we threshold the segmentation area with $\mathbf{e}=0.4$, and the final mask output is defined by $M_{query} = score \geq \mathbf{e}$. With text-promptable 3D querying, surgeons can use natural language to obtain the 3D semantic information of surgical instruments and anatomical structures in real-time.\\

\noindent \textbf{Semantic Region-Aware Optimization.} 
Given a semantic feature map $F$, we introduce a region-based loss to enforce a more accurate and consistent semantic feature for surgical scenes. We perform a specific optimization for each unique semantic area $l$ greater than 1000 pixels within the ground truth $\hat{F}$. Our proposed region-aware smoothness loss is defined as:
\begin{equation}
\mathcal{L}_{RS}=\sum_{l\in \hat{F}}\sum_{(i,j)\in l}\vert F_{ij}-\text{mean}(F_{l})\vert, \ | \ \vert l \vert>1000
\end{equation}
 The region-aware loss ensures that the semantic features within large regions are more consistent and accurate, benefiting both the reconstruction of rendering color and semantic information with clearer object boundaries. Additionally, we apply the L1 loss and total variation (TV) loss $L_{tv}$ to supervise color, depth, and semantic features. Our final loss is presented as follows:
\begin{equation}
    \mathcal{L} = \vert C-\hat{C}\vert + \vert \frac{D-\hat{D}}{D\hat{D}}\vert + \vert F-\hat{F}\vert+ \lambda (\mathcal{L}_{tv}(C)+\mathcal{L}_{tv}(D)+\mathcal{L}_{tv}(F)+\mathcal{L}_{RS}),
\end{equation}
where $\lambda$ is the weight for the TV loss and the region-aware smoothness. $\hat{C}$ is the ground truth color, $\hat{D}$ is the metric depth obtained from pre-trained model~\cite{cui2024endodac}.

%% file: sections/3_experiments.tex
\subsection{Dataset}
\label{sec:dataset}

Our evaluation utilizes two public datasets, CholecSeg8K~\cite{hong2020cholecseg8k} and EndoVis18~\cite{allan20202018}. CholecSeg8K is a comprehensive semantic segmentation dataset derived from the Cholec80~\cite{twinanda2016endonet} dataset for laparoscopic cholecystectomy procedures, containing various annotations. EndoVis18 is a robotic scene segmentation dataset involving surgical instruments and anatomical classes. In our evaluation, we utilize five video sequences (01\_00080, 01\_00240, 01\_15019, 12\_15750, 17\_01803) from CholecSeg8K and two subsets of video sequences (Seq\_5, Seq\_9) from EndoVis18. We split the training and testing images following~\cite{zha2023endosurf} and evaluate the segmentation with mean intersection over union (mIoU). We also evaluate the training time and query speed to compare the model efficiency.

\input{tabs/tab1}
\input{tabs/tab2}
\subsection{Implementation Details}
\label{sec:implementation}

We run all experiments on a single RTX4090 GPU. We utilize Adam optimizer with an initial learning rate of $1.6\times 10^{-3}$. We train CAT-Seg~\cite{cho2024cat} with the CholecSeg8K~\cite{hong2020cholecseg8k} and EndoVis18~\cite{allan20202018}, excluding all sequences in our experiments. For EndoVis18, we use half resolution for efficiency. The weight $\lambda$ is set to $0.01$.
\begin{figure}[!t]
    \centering
    \includegraphics[width=\linewidth]{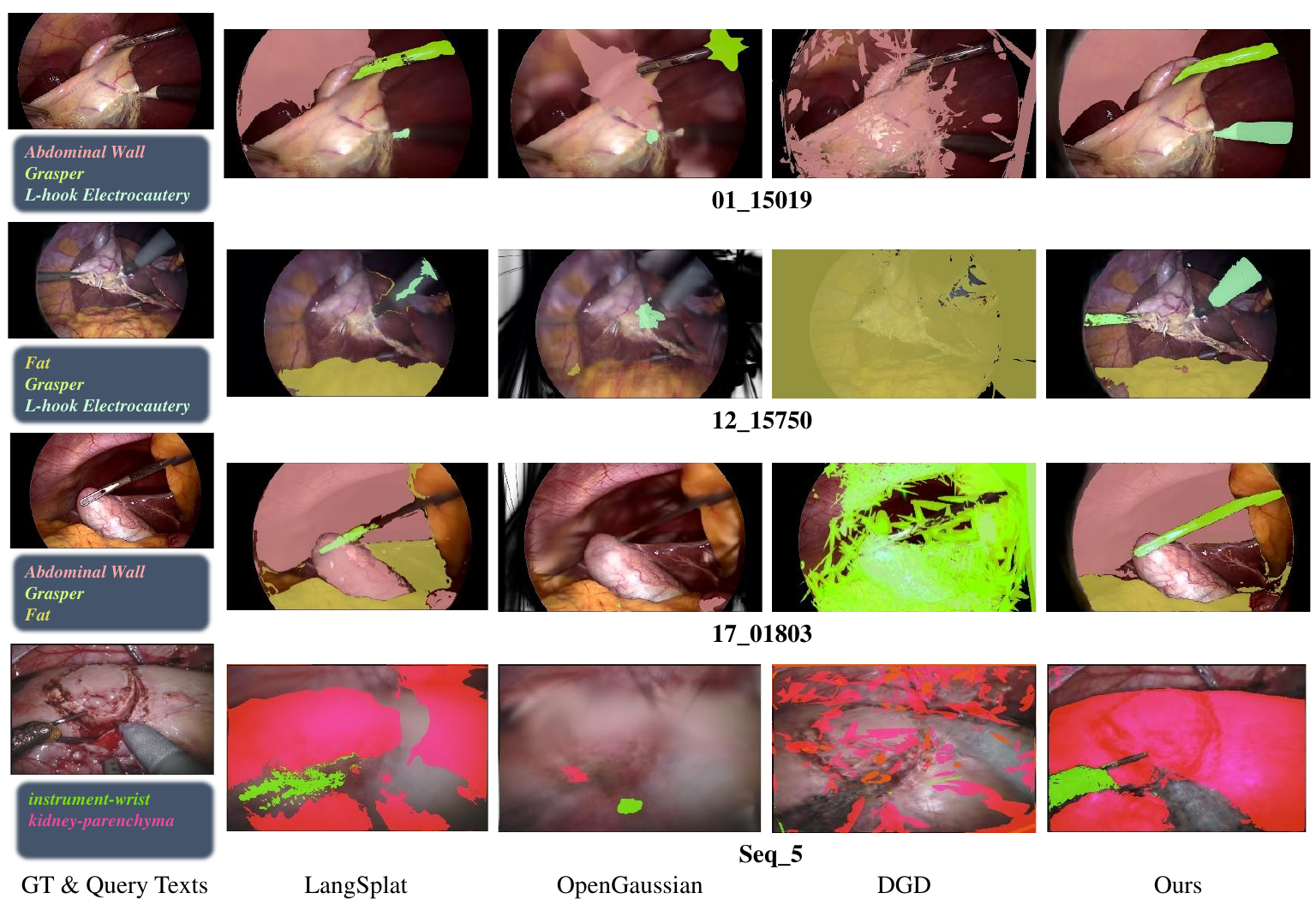}
    \caption{\textbf{Qualitative result on CholecSeg8K~\cite{hong2020cholecseg8k} and EndoVis18~\cite{allan20202018} dataset}. We demonstrate the results of LangSplat~\cite{qin2024langsplat}, OpenGaussian~\cite{wu2024opengaussian}, and ours with CAT-Seg~\cite{cho2024cat} as VLM. Our method shows more consistent and precise segmentation.}
    \label{fig:qualitative}
\end{figure}
\subsection{Results}
We compare our proposed SurgTPGS with state-of-the-art Gaussian Splatting methods for open-vocabulary query: LangSplat~\cite{qin2024langsplat}, OpenGaussian~\cite{wu2024opengaussian}, DGD~\cite{labe2024dgd}. Furthermore, we reimplement LangSplat and OpenGaussian by replacing the vision-language model in their baseline with the SurgVLP~\cite{yuan2023learning} for surgical VQA and CAT-Seg~\cite{cho2024cat}. In addition to the proposed method with CAT-Seg, we also provide the results of our methods in the CLIP and SurgVLP versions. 


As shown in Table~\ref{tab:cholecseg}, our method demonstrates superiority over the baseline approaches on the CholecSeg8K dataset. For example, our method with CAT-Seg achieves a mIoU of 89.82\% for \textit{abdominal wall} in 01\_00080, outperforming LangSplat (CAT-Seg) with a mIoU of 77.51\% and other methods by a large margin. 
Our method also performed excellently in the EndoVis18 dataset in Table~\ref{tab:endovis}. For the kidney-parenchyma segmentation, our method with CAT-Seg achieves a mIoU of 71.98\%, exceeding LangSplat (CAT-Seg), which had 59.17\% and other baselines, while maintaining a high frame rate of 67.5 FPS. Additionally, our method achieves training with only 4 minutes, significantly less than the 8 - 16 minutes required by other methods. This indicates that our method can not only segment surgical objects accurately but also respond quickly to queries, while others fail due to the challenging deformation of surgical scenes. The choice of pre-trained VLM also significantly affects the results. With CAT-Seg, our method can precisely segment the objects in surgical scenes, while with SurgVLP~\cite{yuan2023learning} and CLIP~\cite{radford2021learning} the performance drops due to incorrect mapping from VLM. As shown in Fig~\ref{fig:main}, the quantitative results from both datasets show that our method provides more accurate segmentation with clearer boundaries.

\begin{figure}[!t]
    \centering
    \includegraphics[width=.9\linewidth]{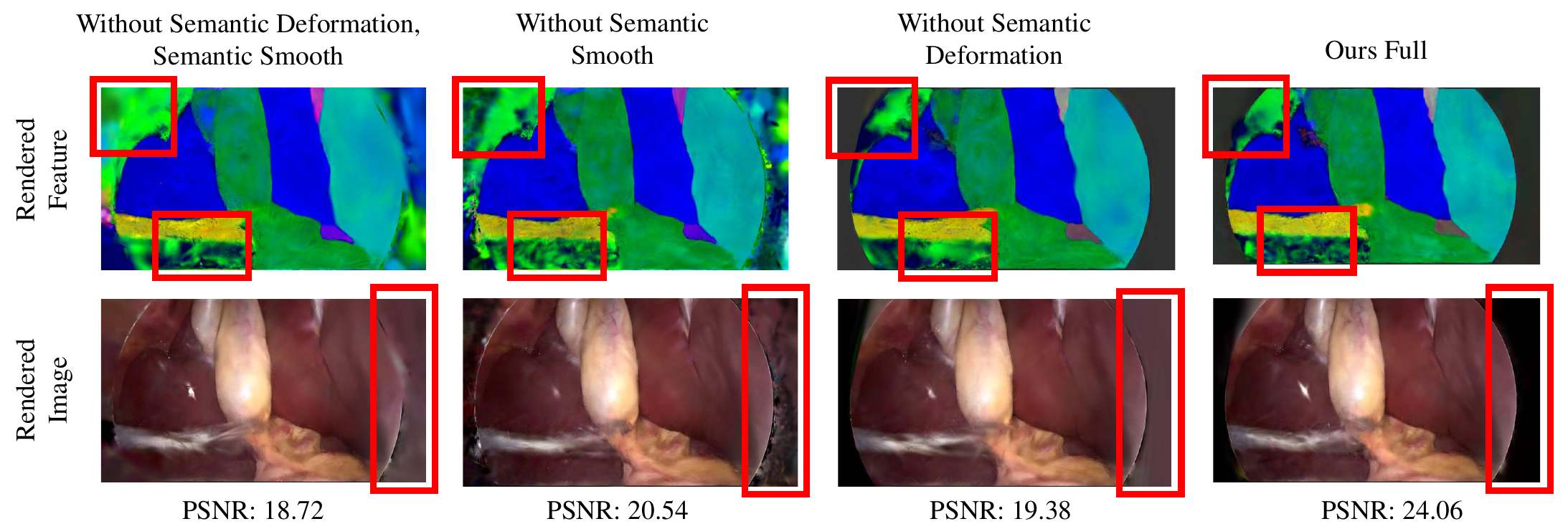}
    \caption{\textbf{Visualization of the ablation results.} Results from our full model present smoother boundaries and clear visualization quality.}
    \label{fig:ablation}
\end{figure}

We conduct the ablation study on the CholecSeg8K dataset, as shown in Table~\ref{tab:ablation} and Fig.~\ref{fig:ablation}. By removing the proposed components, the mIoU and the PSNR decrease, and the visual quality for rendering drops significantly, showing that both components are crucial for accurate semantic feature reconstruction.

\input{tabs/tab3}

%% file: tabs/tab1.tex
\begin{table}[!t]
    \centering
    \caption{
    \textbf{Quantative results on the CholecSeg8K~\cite{hong2020cholecseg8k} dataset}.    
  We compare the mIoU scores (\%), and highlighted the \colorbox{red!40}{first}, \colorbox{orange!50}{second}, and \colorbox{yellow!50}{third}. Our method outperforms state-of-the-art baseline methods.
\label{tab:cholecseg}
    }
    \resizebox{\textwidth}{!}{ 
        \begin{tabular}{@{}l|ccccccccc@{}}
        \toprule
            \multirow{2}{*}{Method}& \multicolumn{1}{c}{\parbox{1.5cm}{\centering 01\_00080}} & \multicolumn{2}{c}{01\_00240} & \multicolumn{2}{c}{01\_15019} & \multicolumn{2}{c}{12\_15750} & \multicolumn{2}{c}{17\_01803}\\
            & Liver  & Liver & Grasper& \parbox{1.8cm}{\centering Abdominal \\ Wall}& Grasper & Fat & \parbox{1.8cm}{\centering L-hook \\Electrocautery}& \parbox{1.8cm}{\centering Abdominal \\Wall} & Grasper\\
            \midrule
            LangSplat\cite{qin2024langsplat} & 64.87 & \colorbox{yellow!50}{57.91} & 4.5& 23.84& 4.13 & 25.63 & 5.97 & \colorbox{yellow!50}{43.38} & 4.13 \\
            LangSplat (SurgVLP\cite{yuan2023learning}) & 65.03 & 57.14 & 4.31& 32.79 & 5.61 & 4.1 & 5.97 & 42.97 & 4.53 \\
            LangSplat (CAT-Seg\cite{cho2024cat}) & \colorbox{orange!50}{77.51} & \colorbox{orange!50}{73.69}& \colorbox{yellow!50}{4.96}  & \colorbox{orange!50}{91.05} & \colorbox{orange!50}{58.29} & \colorbox{red!40}{75.83} & \colorbox{orange!50}{23.81} & \colorbox{orange!50}{72.93} & \colorbox{orange!50}{14.46} \\
            OpenGaussian\cite{wu2024opengaussian} & 2.64 & 0.08 & 0 & 0 & 0 & 0 & \colorbox{yellow!50}{8.05} & 0 & 0 \\
            OpenGaussian (SurgVLP\cite{yuan2023learning}) & 1.76 & 0 & 0 & 0 & 0 & 0.19 & 0 & 13.02 & 2.24 \\
            OpenGaussian (CAT-Seg\cite{cho2024cat}) & 2.21 & 1.9 & \colorbox{orange!50}{12.79} & 14.89 & \colorbox{yellow!50}{12.82} & 1.26 & 7.49 & 0 & 0 \\
            DGD\cite{labe2024dgd} & 13.98 & 42.43 &  3.83 & 9.68 & 2.9 & 14.79 & 0 & 4.09 & 2.67 \\ \midrule
            Ours (CLIP\cite{radford2021learning}) & 64.96 & 57.52 &  4.69& 15.94 & 4.1 & \colorbox{yellow!50}{26.59} & 5.34 & 23.55 & \colorbox{yellow!50}{7.18} \\
            Ours (SurgVLP\cite{yuan2023learning}) & \colorbox{yellow!50}{65.07} & 57.04 & 6.01 & \colorbox{yellow!50}{32.99} & 5.91 & 8.03 & 6.14 & 42.94 & 4.56 \\
            \textbf{Ours (CAT-Seg\cite{cho2024cat})} & \colorbox{red!40}{89.82} & \colorbox{red!40}{79.06} & \colorbox{red!40}{65.78} & \colorbox{red!40}{97.24} & \colorbox{red!40}{68.83} & \colorbox{orange!50}{74.84} & \colorbox{red!40}{73.29} & \colorbox{red!40}{92.12} & \colorbox{red!40}{55.04} \\
            \bottomrule
        \end{tabular}
    }


\end{table}

%% file: tabs/tab2.tex
\begin{table}[!t]
\centering%

    \caption{
    \textbf{Quantative results on the EndoVis18~\cite{allan20202018} dataset}.
  We highlighted \colorbox{red!40}{first}, \colorbox{orange!50}{second}, and \colorbox{yellow!50}{third} values. We compare the mIoU scores (\%) for segmentation, and query speed (FPS), training time for model efficiency. The performance of our proposed method surpasses state-of-the-art baselines.
\label{tab:endovis}
    }
\resizebox{\textwidth}{!}{
        \begin{tabular}{@{}l|ccccccc@{}}
            \toprule
            \multirow{3}{*}{Method}& \multicolumn{2}{c}{\parbox{1.5cm}{\centering Seq\_5}} & \multicolumn{3}{c}{Seq\_9} & \multirow{3}{*}{\parbox{1cm}{\centering Query \\FPS$\uparrow$}}&\multirow{3}{*}{\parbox{1.8cm}{\centering Training \\ Time$\downarrow$}}\\
            & \parbox{1.8cm}{\centering instrument\\-wrist} & \parbox{1.9cm}{\centering kidney\\-parenchyma} & \parbox{1.8cm}{\centering instrument\\-shaft}& \parbox{1.6cm}{\centering instrument\\-wrist}& \parbox{1.8cm}{\centering instrument\\-clasper} &  & \\
            \midrule
            LangSplat\cite{qin2024langsplat} & 5.17 & \colorbox{yellow!50}{59.16} & 16.47 & 7.89 & 6.15 & 67.5 & 8 min \\
            LangSplat (SurgVLP\cite{yuan2023learning}) & 7.36 & 52.94 & 9.87 & 11.72 & 7.02 & 23.8 & 9 min  \\
            LangSplat (CAT-Seg\cite{cho2024cat}) & \colorbox{orange!50}{17.16} & 54.42 & \colorbox{red!40}{38.49} & \colorbox{orange!50}{21.74} & 0.11 & 67.5 & 8 min \\
            OpenGaussian\cite{wu2024opengaussian} & 0 & 1.62 & 0 & 0 & 0 & \colorbox{red!40}{378.9} & 15 min \\
            OpenGaussian (SurgVLP~\cite{yuan2023learning}) & 1.13 & 0.54 & 0 & 0 & 0 & \colorbox{yellow!50}{118.5} & 16 min  \\
            OpenGaussian (CAT-Seg\cite{cho2024cat}) & 0 & 0.06 & \colorbox{orange!50}{29.38} & \colorbox{yellow!50}{12.82} & \colorbox{yellow!50}{14.89} & \colorbox{red!40}{378.9} & 15 min  \\
            DGD\cite{labe2024dgd} & 1.14 & 21.68 & 8.22 & 5.04 & 7.37 & 4.43 & 26 hrs  \\ \midrule
            Ours (CLIP\cite{radford2021learning}) & 5.17 & \colorbox{orange!50}{59.17} & 15.55 & 4.1 & \colorbox{orange!50}{15.94} & 67.5 & \colorbox{red!40}{4 min}  \\
            Ours (SurgVLP\cite{yuan2023learning}) & \colorbox{yellow!50}{11.2} & 21.26 & 25.3 & 4.73 & 3.53 & 23.8 & \colorbox{yellow!50}{5 min} \\
            \textbf{Ours (CAT-Seg\cite{cho2024cat})} & \colorbox{red!40}{43.29} & \colorbox{red!40}{71.98} & \colorbox{yellow!50}{20.79} & \colorbox{red!40}{24.34} & \colorbox{red!40}{38.94} & 67.5 & \colorbox{red!40}{4 min} \\
            \bottomrule
        \end{tabular}
    }


\end{table}

%% file: tabs/tab3.tex
\begin{table}[!t]
    \centering%
    \caption{\textbf{Ablation results\label{tab:ablation} on 01\_00240 of CholecSeg8K~\cite{hong2020cholecseg8k}}. We compare the mIoU and the PSNR of rendering results. The \colorbox{red!40}{first} and \colorbox{orange!50}{second} are highlighted.}

    \resizebox{0.7\textwidth}{!}{
        \begin{tabular}{cc|ccc}
        \toprule
        
        \multirow{2}{*}{\parbox{3.5cm}{\centering Semantic-Aware \\Deformation Tracking}} 
        & \multirow{2}{*}{\parbox{2.5cm}{\centering Region-Aware Smoothness}} 
        & \multicolumn{2}{c}{mIoU$\uparrow$} 
        & \multirow{2}{*}{\parbox{1.5cm}{\centering PSNR$\uparrow$}}\\
        & & \parbox{1.5cm}{\centering Liver} 
        & \parbox{1.5cm}{\centering Grasper} \\ 
        \midrule
        \XSolidBrush  & \XSolidBrush & \colorbox{red!40}{79.43} & 64.02 & 18.72 \\
        \Checkmark & \XSolidBrush & 78.96 & 62.79 & \colorbox{orange!50}{20.54} \\
        \XSolidBrush & \Checkmark & 78.85 & \colorbox{orange!50}{64.78} & 19.38 \\
        \Checkmark & \Checkmark & \colorbox{orange!50}{79.06} & \colorbox{red!40}{65.78} & \colorbox{red!40}{24.06} \\
        \bottomrule\\
        \end{tabular}
        }

\end{table}

%% file: sections/4_conclusion.tex
In conclusion, SurgTPGS represents a significant advancement in understanding 3D surgical scenes. By integrating a 3D semantics feature learning strategy, semantic-aware deformation tracking, and semantic region-aware optimization, our method addresses the limitations of existing methods and outperforms state-of-the-art techniques on real-world surgical datasets in terms of text-promptable segmentation accuracy, maintaining efficient training time and query speed. SurgTPGS can revolutionize robot-assisted surgery by enabling robots to better understand and interact with the surgical environment, leading to more precise and safer robot-assisted surgery procedures.